\documentclass[11pt,twoside,a4paper]{article}
\usepackage{graphicx}
\usepackage{subcaption}
\usepackage{amsmath}
\usepackage{amssymb}
\usepackage{xcolor}
\usepackage[normalem]{ulem}

\newcommand{\e}[1]{\mathrm{e}^{#1}}

\title{An interferometric method for determining the losses of spatially multi-mode nonlinear waveguides based on second harmonic generation.}
\author{Matteo Santandrea, Michael Stefszky, Gana\"el Roeland, Christine Silberhorn}
\date{$^1$ Integrated Quantum Optics, Paderborn University, Warburger Stra{\ss}e 100, 33098 Paderborn, Germany\\
$^2$ Laboratoire Kastler Brossel, Sorbonne Universit\'e,  CNRS, ENS-PSL Research University, Coll\`ege de France, 4 place Jussieu, F-75252 Paris, France\\
$^*$ Corresponding author: matteo.santandrea@upb.de}

\begin{document}
\maketitle
\begin{abstract}
The characterisation of loss in optical waveguides is essential in understanding the performance of these devices and their limitations. Whilst interferometric-based methods generally provide the best results for low-loss waveguides, they are almost exclusively used to provide characterization in cases where the waveguide is spatially single-mode.  Here, we introduce a Fabry-P\'erot-based scheme to estimate the losses of a nonlinear (birefringent or quasi-phase matched) waveguide at a wavelength where it is multi-mode. The method involves measuring the generated second harmonic power as the pump wavelength is scanned over the phasematching region. Furthermore, it is shown that this method allows one to infer the losses of different second harmonic spatial modes by scanning the pump field over the separated phase matching spectra. By fitting the measured phasematching spectra from a titanium indiffused lithium niobate waveguide sample to the model presented in this paper, it is shown that one can estimate the second harmonic losses of a single spatial-mode, at wavelengths where the waveguides are spatially multi-mode.
\end{abstract}

\section{Introduction}

Optical waveguides have enabled the expansion of optical networks in a very short time. Naturally, this technology is advancing in order to include, for example, high power, high efficiency and/or quantum applications \cite{Stefszky17.PRA,Stefszky17Comb.A}. For the most demanding applications, such as squeezing in fibre networks \cite{Kaiser16.O} or on-chip entanglement \cite{Barral17.PRA}, the losses of the waveguide are of critical importance. Therefore, the reliable characterisation of these losses is a critical issue.

A number of methods for loss characterization and variants of these methods exist. These methods can be categorised into a few broad schemes: cut-back methods, fluorescence/scatter imaging, resonance techniques and optical transmission measures. These various methods perform differently under given circumstances. Interferometric methods tend to have greater precision as the losses decrease and so are more suited for characterisation of low loss waveguides \cite{Regener85.APB,Park95.JAP,Clark90.OL}.

% For example, cut-back and transmission methods tend to perform poorly when the waveguide losses are low. This is primarily due to the fact that the uncertainty in the coupling of the field to the waveguide can dominate the waveguide losses themselves for very low-loss waveguides. This class of characterization methods is generally not suitable for state-of-the-art waveguides, such as titanium indiffused waveguides, which now attain losses as low as 0.02dB/cm \cite{Luo15.NJP}. In contrast, interferometric methods tend to have greater precision as the losses decrease and so are more suited for characterisation of low loss waveguides \cite{Regener85.APB}. These methods make use of the fact that the transmitted (and/or reflected \cite{Park95.JAP,Clark90.OL}) intensities are highly dependent on the end-face reflectivities and internal loss of the resonator.  Whilst one can use optical coatings on the waveguide end facets to enhance the resonance effect\cite{Stefszky18.JO}, it is typically sufficient to simply use the Fresnel reflections in the so-called low-finesse Fabry-Perot method to reliably characterise the losses in low-loss waveguides \cite{Regener85.APB}.

Unfortunately, such resonance-based methods are generally unsuitable in waveguides that are spatially multi-mode for the probe field. This is due to the fact that it is experimentally very difficult to couple light into the waveguide such that only a single propagation-mode of the waveguide is excited. These different spatial modes have disparate dispersion properties, leading to different free spectral ranges (FSRs) for the various spatial modes. The resulting transmitted power will consist of multiple resonance conditions with unknown magnitude and phase, generally making the problem intractable. Under certain conditions the losses can still be obtained from such a measurement, but this requires the fulfillment of a number of conditions which are, in general, not satisfied \cite{DeRossi05.JAP}.

For this reason, devices implementing multi-colour processes, such as difference or sum-frequency generation, typically have their losses characterised at the longest wavelength, where the waveguide is single-mode. This value is often used to estimate or bound the losses at shorter wavelengths. However, one cannot know \textit{a priori} the exact relationship between the losses at different wavelengths. This is problematic when the losses at the shorter wavelengths are critical, for example in frequency converters that aim to produce a field close to the transparency cut-off region of a particular material \cite{Ruetz16.APB}.

%For many applications the losses at the shorter wavelength/s are critical. For example in frequency converters that aim to produce a field close to the transparency cut-off region of a particular material \cite{Ruetz16.APB} or any device that aims to resonate the shorter wavelength fields. Therefore, methods to non-destructively characterise the propagation losses at wavelengths where the waveguides are multi-mode are required.

Here we present a method for loss characterisation in such systems by measuring the phase matching spectrum of the second harmonic (SH) process as the pump (fundamental) wavelength is varied over the phasematching spectrum. The resulting phase matching spectrum is compared to theory in order to estimate the losses of the second harmonic field. Additionally, this method allows one to probe the second harmonic losses for \textit{the spatial mode of one's choosing}. The approach is quite general and can be applied to both birefringent and quasi-phase matched systems. One could extend the theory to other processes such as sum frequency generation and type II second-harmonic generation processes.

%This method also allows for selective measurement of the losses of the different spatial modes, as these spatial modes have unique dispersion properties and therefore individual phasematching spectra. A simple modification of this technique should allow for it to also be applied to many sum and difference frequency generation schemes. Furthermore, this allows one to probe the second harmonic losses for \textit{the spatial mode of one's choosing} by centering the wavelength of the fundamental pump field to the various phasematching profiles of the second harmonic modes. The approach is quite general and in principle could be extended to other processes such as sum frequency generation and type II second-harmonic generation processes.

\section{Measurement Strategy and Theory}

In the standard low-finesse Fabry-Perot loss measurement the power transmitted through a waveguide is recorded when scanning a probe field over wavelengths where the system is single-mode \cite{Regener85.APB}. Given that one knows the reflectance of the end facets to a high precision, the interference effects observed in the transmitted power can be used to determine the losses inside the resonator at the same wavelength as the probe field. 

The general strategy employed in the method presented here is that, in addition to first determining the losses at the fundamental wavelength using the standard method, we also measure the interferometric fringing that one observes in the generated second harmonic field when scanning the pump over the wavelengths where phasematching occurs. One can then fit the obtained phasematching spectrum to a model of this system and gain information about the losses of the second harmonic field. A single spatial-mode for the second harmonic field is guaranteed due to the fact that the single-mode pump field is phasematched to only a single second harmonic spatial-mode over the wavelength region of interest. The unique dispersion properties of different spatial modes generally ensures that this is the case.

The system is modeled using an extension of the second harmonic generation theory presented by Berger \cite{Berger97.JOSAB}. In this method, the internal second harmonic fields are first described and thereafter solved simultaneously in order to find a self-consistent cavity solution. In order to arrive at an analytic expression it is assumed that the pump field, at the fundamental frequency, is not depleted by the nonlinear process. This assumption is trivial to establish experimentally by correctly choosing the power in the pump field. It may be possible to remove this restriction by considering a numerically based iterative approach \cite{Fujimura96.JLT}. However, this will further complicate the treatment and will not provide an analytic expression.

The circulating fundamental field \textit{amplitude} travelling in the cavity in the forward direction $E^f_{\omega}(0)$ is given by the usual Fabry-Perot resonance condition
\begin{equation}
E^f_{\omega}(0) = E_{in}\frac{\tau_{\omega, 0} }{1-\rho_{\omega, 0} \rho_{\omega, L} \cdot \e{-i 2 k_\omega L}\cdot \e{-\alpha_\omega L}}\label{eq:fundamentalfringing}
\end{equation}
where $k_\omega$ $[m^{-1}]$ is the wavevector of the fundamental field,  $\alpha_{\omega}$ $[m^{-1}]$ are the intensity losses for the fundamental field, $\rho_{\omega, 0/L}$  is the complex reflectivity for the input/output facet at $\omega$ and $\tau_{\omega,0}$ is the complex transmission of the input facet at $\omega$. Energy conservation ensures that  $|\rho_{\omega}|^2 + |\tau_\omega|^2 = 1$. Note that from these definitions one can also express the circulating fundamental field \textit{amplitude} travelling in the backwards direction, $E_\omega^b(L) =\rho_{\omega, L} E_\omega \e{ik_\omega L}$.

With the non-pump depletion approximation, the generated second harmonic field \textit{amplitude} can be calculated from \cite{Fujimura96.JLT} as 
\begin{equation}
\frac{\mathrm{d} E_{2\omega}}{d z} =  i \gamma [E_{\omega}\e{-\alpha_\omega z/2}]^2 \e{i\Delta k z} - \frac{\alpha_{2\omega}}{2}E_{2\omega}, \label{eq:nlODE}
\end{equation}
where $E_{\omega/2\omega}(z)$ is the fundamental/second harmonic field amplitude at position $z$,  $\alpha_{2\omega}$ $[m^{-1}]$ represents the (intensity) losses of the second harmonic field, $\gamma$ $[m/V]$ is the nonlinear coupling coefficient determining the strength of the nonlinear process, the wave vector mismatch between the fundamental and second harmonic field is defined by $\Delta k = 2k_\omega-k_{2\omega} + k_{QPM}$ $[m^{-1}]$, and $k_{2 \omega}$ $[m^{-1}]$ is the wave vector of the second harmonic field. The term $k_{QPM}=2\pi/\Lambda_0$ is required only when analysing periodically poled structures, with period $\Lambda_0$. Note that  the effect of losses in the fundamental field in eq. \eqref{eq:nlODE} have been included by considering a spatially dependent fundamental amplitude in the form $E_{\omega}\e{-\alpha_\omega z/2}$.

Integration of eq. \ref{eq:nlODE} with initial conditions $\left(E_{\omega}(z_0),E_{2\omega}(z_0)\right)$  over a crystal length $L$ yields the component of the second harmonic amplitude after passing through the length $z$ due to the nonlinear interaction 
\begin{eqnarray}
E_{2\omega}(z) &=& SH_z\left(E_\omega(z_0), E_{2\omega}(z_0)  \right),\\
&=& 2i\gamma \frac{   \e{(\alpha_{2\omega}-2\alpha_\omega+2i\Delta k)z/2}-1 }{ \alpha_{2\omega}  -2\alpha_\omega+ 2 i \Delta k}E_\omega^2\e{-\alpha_{2\omega}z/2} + {E}_{2\omega}\e{-\alpha_{2\omega}z/2}.
\end{eqnarray}

\begin{figure}[!ht]
	\centering
	\includegraphics[width = 10cm]{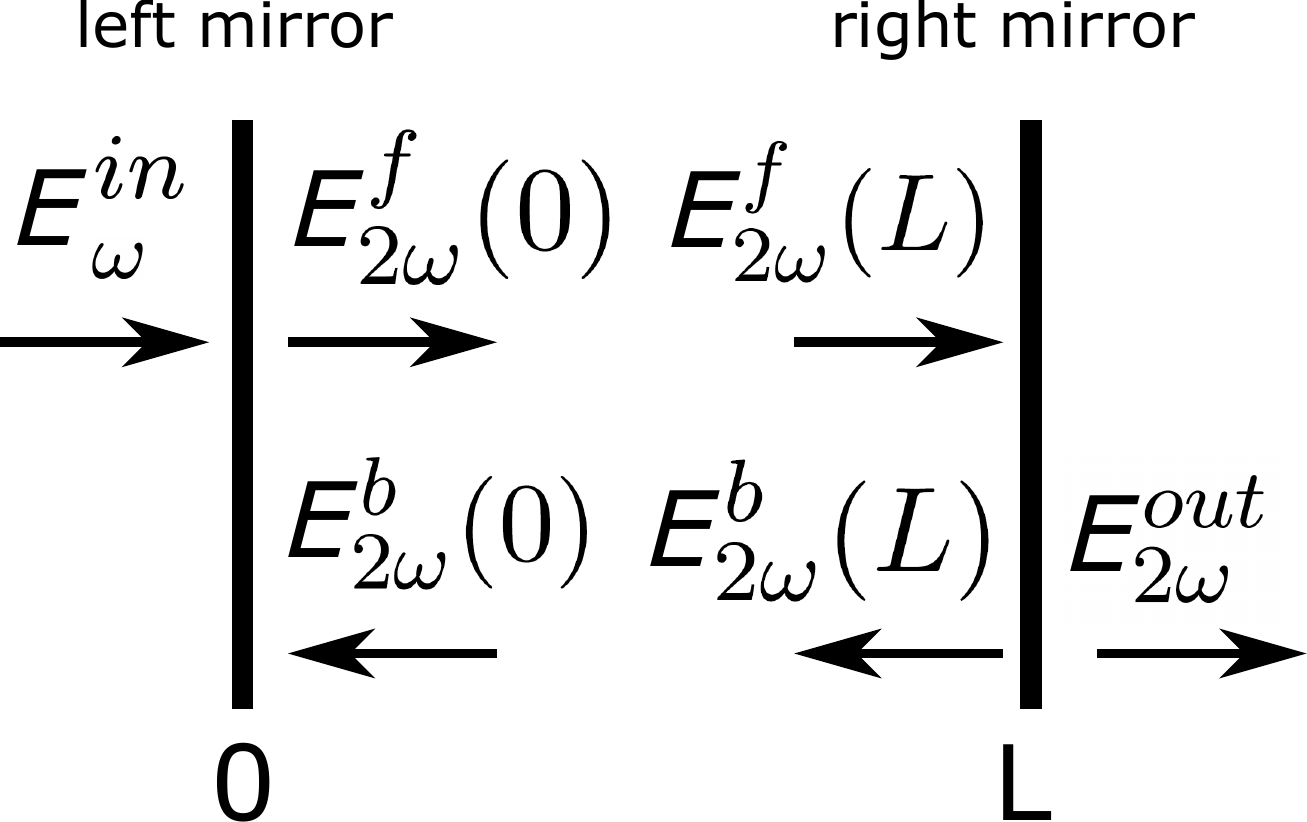}
	\caption{Sketch detailing the forward and backward propagating waves used for the theoretical treatment of the waveguide resonator.}
	\label{Spectra}
\end{figure}

To derive an expression for the circulating second harmonic field amplitude, one defines the second harmonic field \textit{amplitudes} travelling in the forward direction at the left and right sides of the sample, $E^f_{2\omega}(0)$ and $E^f_{2\omega}(L)$, and in the backwards direction at the left and right sides of the sample, $E^b_{2\omega}(0)$ and $E^b_{2\omega}(L)$, respectively, as illustrated in Figure \ref{Spectra}.
The relation between these four amplitudes can be described by the following system:
\begin{eqnarray}
E^f_{2\omega}(L) &=& SH_L\left(E_\omega^f(0), E^f_{2\omega}(0)\right) e^{i k_{2\omega} L},\\
E^b_{2 \omega}(L) &=& \rho_{2\omega,0}	E^f_{2 \omega},\\
E^b_{2 \omega}(0) &=& SH_L\left(E_\omega^b(L), E^b_{2\omega}(L) \right) e^{i k_{2\omega} L},\\
E^f_{2\omega}(0) &=&  \rho_{2\omega,L}	E^b_{2 \omega},
\end{eqnarray}
where $E^b_\omega(L) = \rho_{\omega, L}E^f_\omega(0)\e{-i k_\omega L}\e{-\alpha_\omega L/2}$.
The total circulating second harmonic field at steady-state can be found by simultaneously solving these equations, thereby ensuring self-consistency of the SH field amplitude.

Solving this set of equations, propagating through the right side mirror in order to find the second harmonic field exiting the cavity and substituting \eqref{eq:fundamentalfringing} we find
%\begin{align}
%E_{2\omega}^{out} =&\tau_{2\omega,L}E_{2\omega,L}^f	\nonumber\\
%=& \tau_{2\omega,L}E_\omega^f(0)^2\gamma \times\nonumber\\
%&\frac{   \e{L/2 (\alpha_{2\omega} +i 2\Delta k )}-1  }{2\Delta k - i \alpha_{2\omega}}\times \nonumber\\
%&\frac{1}{1-\rho_{2\omega,0}\rho_{2\omega,L} \e{-i 2k_{2\omega} L-\alpha_{2\omega}L}} \times\nonumber\\
%&\left(1+\rho_{2\omega,0}\rho_{\omega,L}^2 \e{-i k_{2\omega}L}\e{ -i 2k_\omega L}\e{- \alpha_{2\omega}L/2  -\alpha_\omega L }\right) \e{-\alpha_{2\omega} L - i k_{2\omega L}}
%\end{align}
the output second harmonic field amplitude as 
\begin{align}
E_{2\omega}^{out} =&\tau_{2\omega,L}E_{2\omega,L}^f	\nonumber\\
=& \tau_{2\omega,L}\gamma   (E_{\omega}^{in})^2\frac{\tau_{\omega,0}^2}{(1-\rho_{\omega, 0}\rho_{\omega, L}\e{-2i k_\omega L - \alpha_\omega L})^2}\times\nonumber\\
%&\left[2i \frac{ \e{L/2 (\alpha_{2\omega}  - 2\alpha_\omega + i 2\Delta k )}-1  }{\alpha_{2\omega} -2\alpha_\omega +i 2\Delta k }\right]\times \nonumber\\
&i L \mathrm{sinc}\left(\frac{ (\Delta k - i \alpha_{2\omega}/2 +i \alpha_\omega)L}{2}\right)\e{\frac{ (\Delta k - i \alpha_{2\omega}/2 +i \alpha_\omega)L}{2}}\times \nonumber \\
&\frac{1}{1-\rho_{2\omega,0}\rho_{2\omega,L} \e{-i 2k_{2\omega} L-\alpha_{2\omega}L}} \times\nonumber\\
&\left(1+\rho_{2\omega,0}\rho_{\omega,L}^2 \e{-i k_{2\omega}L}\e{ -i 2k_\omega L}\e{- \alpha_{2\omega}L/2  -\alpha_\omega L }\right) \e{-\alpha_{2\omega} L/2 - i k_{2\omega} L}\label{eq:sh_spectrum}.
\end{align}

This equation is split into four terms in order to highlight the factors that contribute to the observed interference fringes, as noted by Berger \cite{Berger97.JOSAB}. The first term represents the Fabry-Perot interference of the fundamental field;  the second term is the spectrum of the second harmonic signal generated in a single pass; the third term is the Fabry-Perot interference of the second harmonic field and the final term represents the phase mismatch between the nonlinear polarization and the second harmonic field over half of a cavity round trip, or equivalently, the phase between the forward and backwards propagating second harmonic waves.

The second harmonic power exiting the system when pumped at wavelength $\lambda$ is then given by  squaring the field (\ref{eq:sh_spectrum}), $I_{2\omega}(\lambda) = |E_{2\omega}^{out}(\lambda)|^2$.
 
The profile of $I_{2\omega}(\lambda)$ depends on the complex facet reflectivities $\rho=|\rho|\e{i \phi} $ at $z=0$ and $z=L$, the fundamental and SH losses $\alpha_{\omega/2\omega}$ and on the cavity length $L$. Qualitatively, one can observe that these parameters affect the shape of $I_{2\omega}(\lambda)$ in different ways: the length $L$ of the sample affects the width of the spectrum and the free spectral range (FSR) of the primary frequency component of the fringing, the contrast of the fringes depends on the magnitude of both the fundamental and second harmonic losses and the complicated internal structure of the fringing is dependent on the facet reflectivities and the crystal length. 
In the following section it is shown that it is possible to find an optimized fit to these free variables, thereby providing an estimate of the value of $\alpha_{2\omega}$.

\section{Fitting Procedure}

The fit of the theory to the measured data is undertaken in steps in order to constrain the range of some of the parameters to physically acceptable values. First, both the model $I_{2\omega}(\lambda)$ and the measured data $I_{meas}(\lambda)$ are normalized to have unitary maximum intensity. Next, the loss of the fundamental field $\alpha_\omega$ is fixed to the value measured using the standard low-finesse loss technique \cite{Regener85.APB}. This measurement is performed scanning the fundamental field over wavelengths slightly shifted away from phasematching  so that the second harmonic process does not influence the measurement. Next, the length $L_0$ of the sample is retrieved from the free spectral range of the fundamental field. In particular, by Fourier transforming $I_{meas}(\lambda)$, the length $L_0$ is estimated from the FSR or the primary frequency components, corresponding to the interference of the fundamental field \eqref{eq:fundamentalfringing}. Subsequently, the central phasematching wavelength $\lambda_{pm}$ is estimated from the data using a weighted average of the recorded wavelengths, where the second harmonic spectral intensity is used as weights. From $\lambda_{pm}$, the poling period $\Lambda_0$ that best centres the phasematching is chosen. 

After determining the center values of these parameters, the theoretical phasematching spectrum $I_{2\omega}(\lambda)$ is then fitted to the measured data $I_{meas}(\lambda)$. 
As there are a total of 9 free parameters to be optimised (four reflectivity amplitudes $\left|\rho\right|_{\omega/2\omega, 0/L}$, four reflectivity phases $\phi_{\omega/2\omega, 0/L}$ and $\alpha_{2\omega}$), the fit of these quantities is performed in two steps. The phases $\phi$ are first optimised assuming $\alpha_{2\omega} = \alpha_\omega$ and $\left|\rho\right|_{\omega/2\omega, 0/L}$ as given by the corresponding Fresnel equations. 
The optimal $\phi$ are used as initial parameters in the second step of the fit, where the model $I_{2\omega}(\lambda)$ is fitted again to the measured data. At this stage, the length $L$, the poling period $\Lambda_0$, the second harmonic losses $\alpha_{2\omega}$, the modules and phases of the facet reflectivities $\rho_{\omega/2\omega, 0/L}$ are considered as fitting parameters.  
The length $L$ is constrained to a 500$\mu$m range around $L_0$, the poling period $\Lambda_0$ is constrained to be within 1\% of $\Lambda_0$, while the phases retrieved in the first step of the optimisation are used as initial parameters for the fitting algorithm.

The fitting routine solves a nonlinear least square minimisation problem using the Trust Region Reflective algorithm that minimizes the mean squared error (MSE) between the model $I_{2\omega}(\lambda)$ and the data $I_{meas}(\lambda)$. 
Due to the complexity of the model, the initial values for the reflectivities of the facets and $\alpha_{2\omega}$ are initialised with random weights and the minimisation is repeated 10 times to find the best set of parameters. To obtain physically meaningful results, we bound the parameters of the fit during the minimisation. In particular, the phases of $\rho$ are constrained between [0,2$\pi$] and the reflectivities are permitted to vary by a few percent from the calculated values obtained by the Fresnel equation. Moreover, as some measured spectra showed asymmetries attributable to waveguide inhomogeneities \cite{Santandrea19.NJP}, only the central lobe of the second harmonic spectrum was used during the fit.

Note that the length and the poling period are allowed to vary slightly in this fit in order allow some flexibility, required due to phasematching distortions in the measured data. Furthermore, the mirror reflectivities are treated as complex numbers in order to account for an unknown phase shift on reflection at the end facets of the sample. This phase shift can be extended in order to include the unknown phase shifts present in a quasi-phase matched sample. In such samples the length of the first and final domains are generally unknown and will impart an unknown phase shift on the two fields, which can be absorbed by the phase term in the complex reflectivities.

As a final note, the model presented in \eqref{eq:sh_spectrum} requires the refractive indices of both the fundamental and second harmonic fields as the fundamental pump field is varied - the Sellmeier equation. For the titanium-indiffused lithium niobate waveguides investigated here these dispersion relations have been calculated using a finite element solver written in Python implementing the model described in \cite{Strake1988}. This model provides a very accurate description of the dispersion, with a predicted poling period within 0.25$\mu$m of the nominal one (1\% error). In contrast, the bulk model for lithium niobate crystals \cite{Jundt1997} predicts poling periods 1.3$\mu$m away from the nominal ones (8\% error). 

\section{Results}
We apply the described measurement technique in order to retrieve the losses of a 31.2mm long 7$\mu$m-wide titanium indiffused waveguide quasi-phase matched (with a 16.8 $\mu$m poling period) for type 0 second harmonic generation in the TM00 spatial mode when pumped with a fundamental field at 1525nm. This system also supports second harmonic generation in the TM01 mode at around 1480nm. The losses $\alpha_{\omega}$ of each of these phasematching processes at the fundamental wavelength are first found slightly off phasematching. At around 1525nm the fundamental field losses were found to be 0.21 $\pm$ 0.04 dB/cm and at around 1481nm the losses were found to be 0.24 $\pm$ 0.07 dB/cm.

\begin{figure}[tbp]
\centering
\includegraphics[width =0.8 \textwidth]{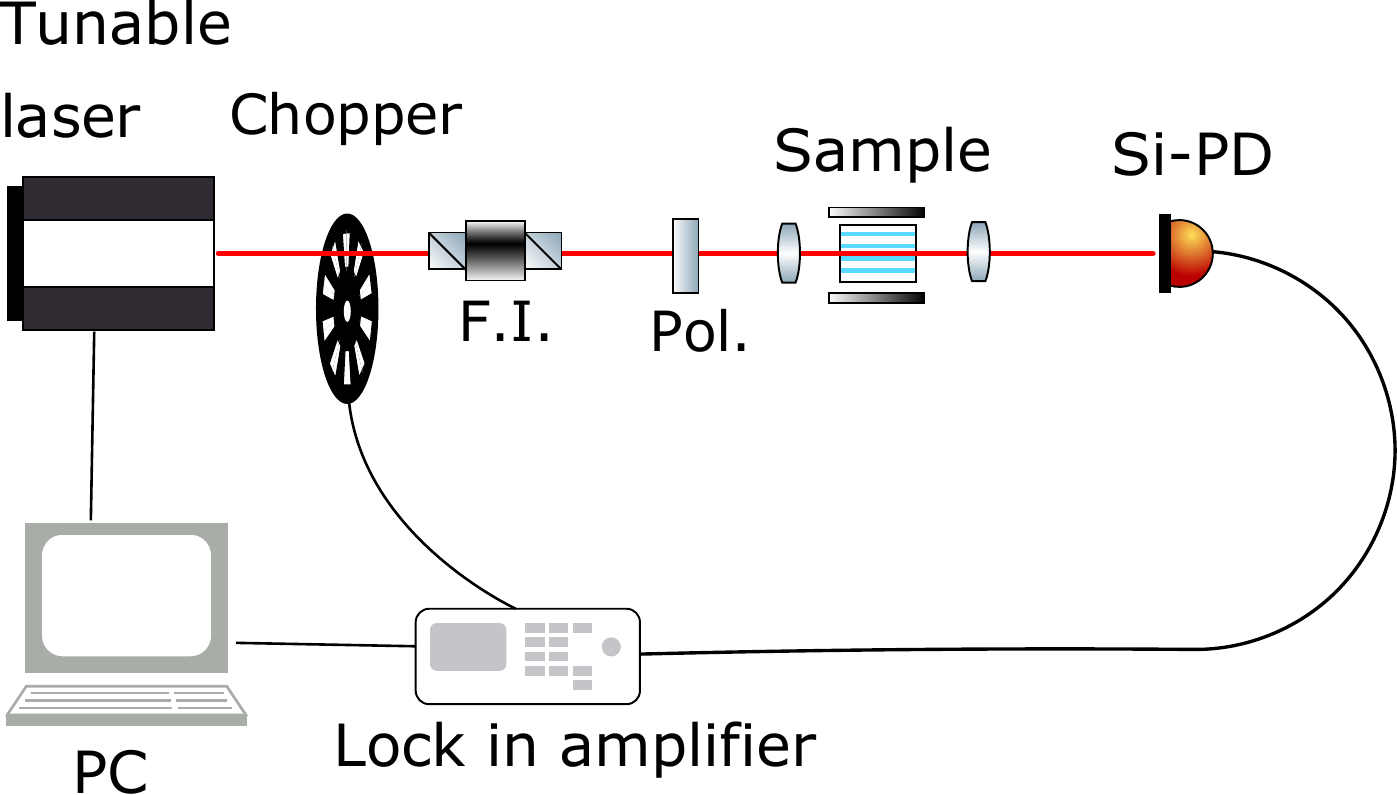}
\caption{Setup for the measurement of the second harmonic. The light from an IR laser tunable in the range 1460nm-1640nm (EXFO TUNICS) passes through a chopper used in conjuction with a lock-in amplifier to enhance the second harmonic readout. The IR field then passes through a Faraday isolator (F.I.) that suppresses any backreflection from the sample. A polariser is used in front of the sample to set the input polarisation of the fundamental field. Anti-reflection (AR) coated 8mm focal length aspheric lenses are used for the in- and out-coupling. Finally, the second harmonic light is measured via a silicon photodiode connected to a lock-in amplifier.}
\label{fig:meas_setup}
\end{figure}

\begin{figure}
\centering
\begin{subfigure}{0.6\linewidth}
\includegraphics[width = \textwidth]{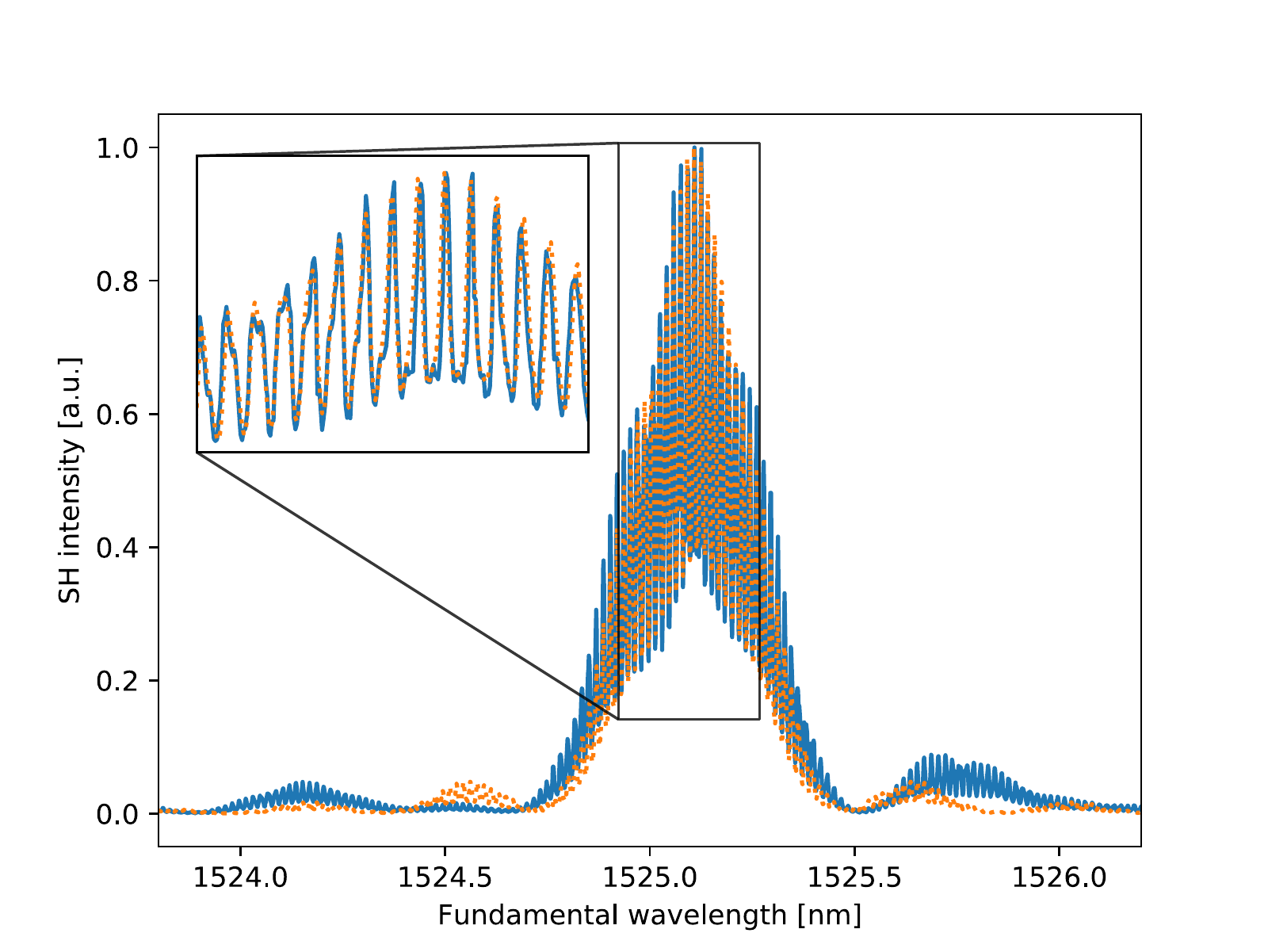}
 \caption{TM00}
\label{tem00}
\end{subfigure}\\	
\begin{subfigure}{0.6\linewidth}
	\includegraphics[width = \textwidth]{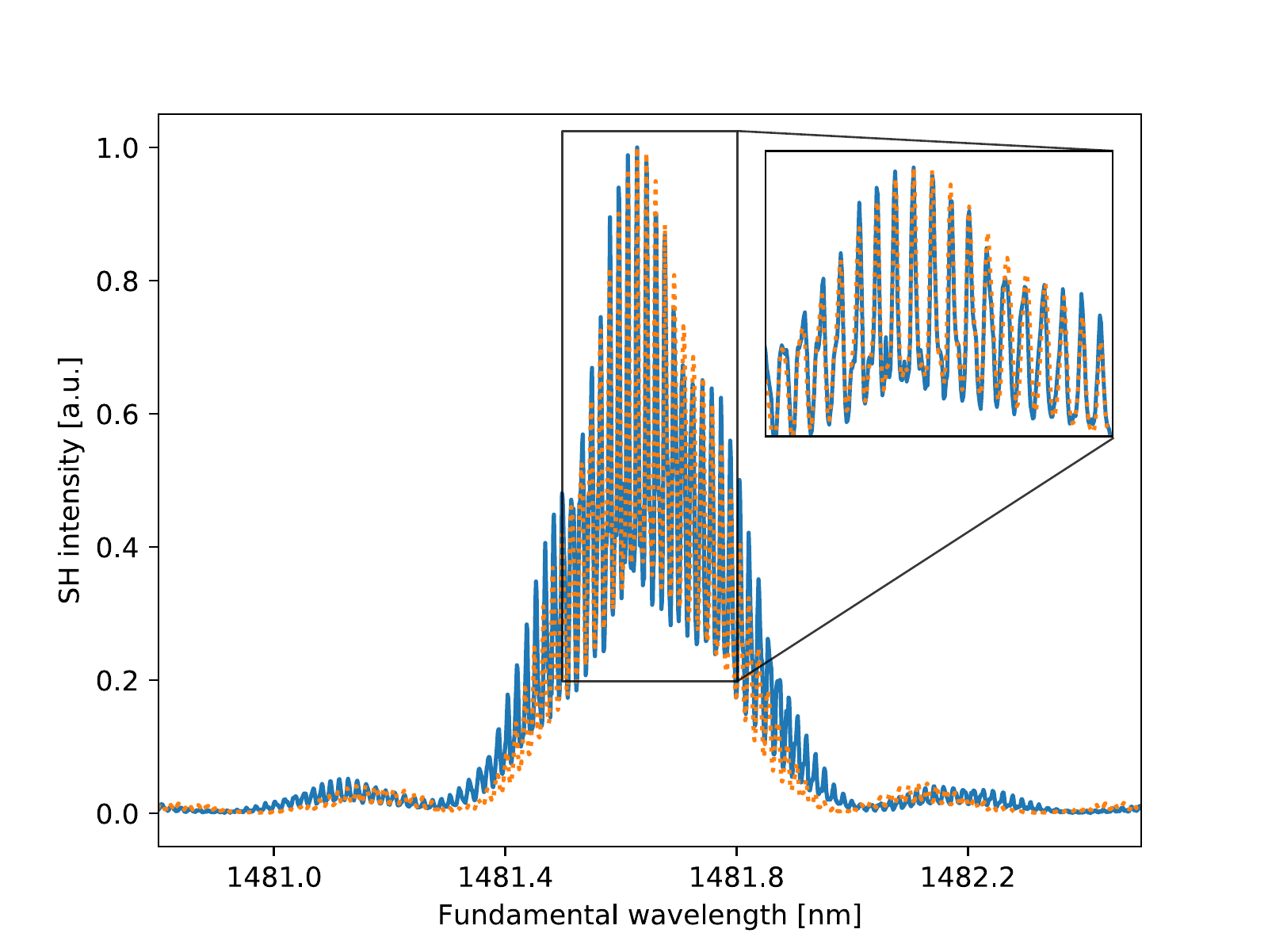}
		\caption{TM01}
	\label{tem10}
\end{subfigure}	
\caption{Measured (blue line) and theoretical fit (orange dotted) for the TM00 and TM01 second harmonic mode phasematching spectra.}
\label{fig:fitting_modes}
\end{figure}

Subsequently, the phasematching spectra of the second harmonic field are recorded as the fundamental wavelength is scanned over the phasematching profile for the TM00 and TM01 second harmonic modes, using the setup shown in Figure \ref{fig:meas_setup}. The measured phasematching spectra and the fits found using the procedure described in the previous section are illustrated in Figure \ref{tem00}.

An excellent qualitative fit between the measured profile and the theory is observed. The frequency of the fringing and the envelope of this central region overlap well. In particular, the insets show zoomed-in regions of the fits that highlight the fact that even the highly complex structure of the interferences is reproduced by the theory. It can be seen, however, that the presence of waveguide imperfections affects the fit of the ``side lobes'' of the profile.  The minimisation routine results in losses of 1.2$\pm$ 0.2 dB/cm for the TM00 second harmonic mode and losses of 1.3$\pm$ 0.1 dB/cm for the TM01 mode, where the errorbar have been derived considering a 1\% variation of the MSE.

\begin{figure}[btp]
\centering
\includegraphics[width = 10cm]{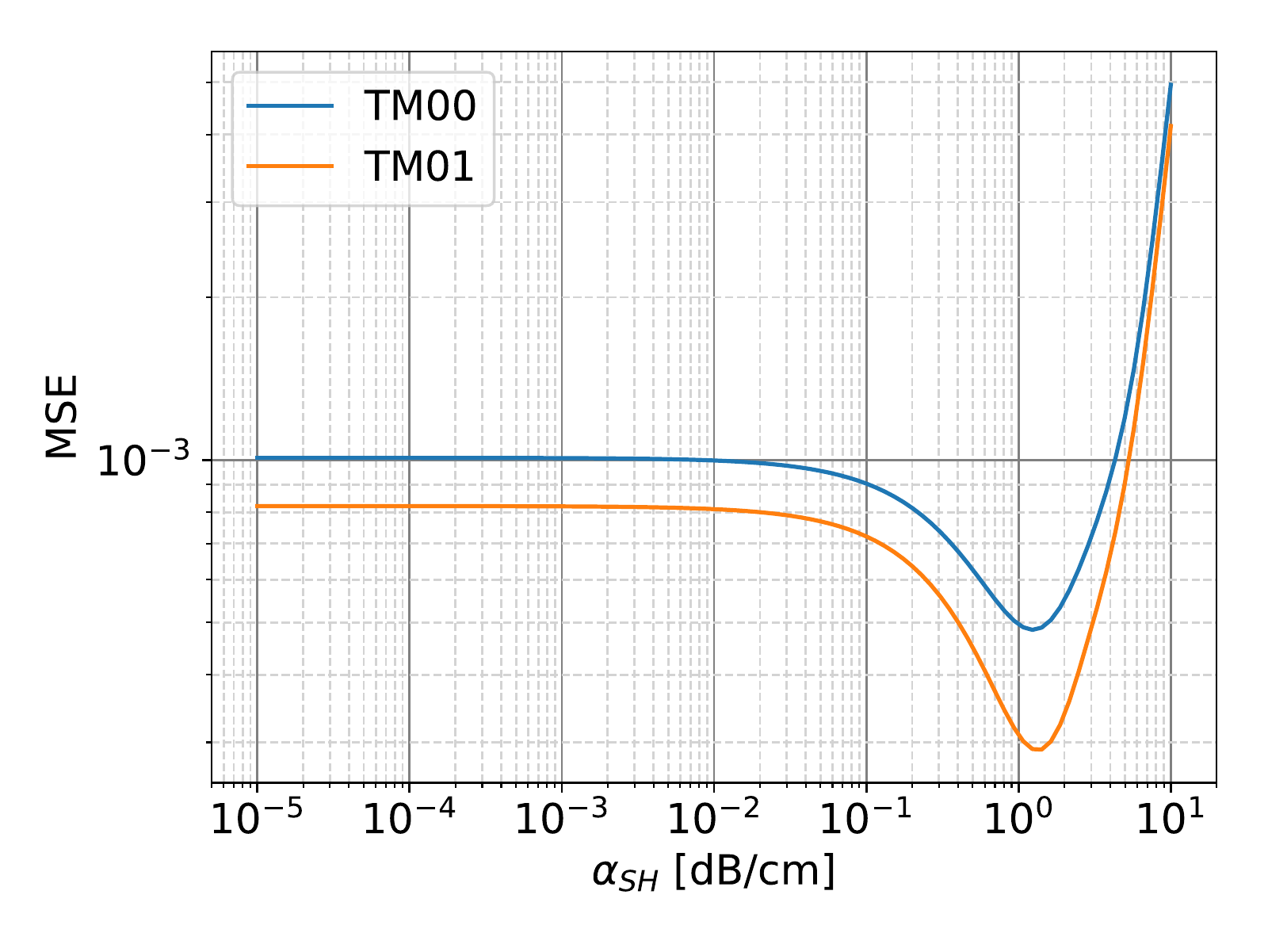}
\caption{Mean squared error of the sum squared residuals between the model and the data for the TM00 and TM01 second harmonic modes.}
\label{img:MSE}
\end{figure}

In order to check the validity of the fit, in particular the performance of the chosen minimisation routine, we also show the variation of the MSE for both of these fits as the second harmonic losses are varied, holding all other parameters constant. The MSE's found in this way are illustrated in Figure \ref{img:MSE}.

The fitting technique employed here is highly sensitive to the shape of the MSE as a function of $\alpha_{2\omega}$. In particular, the MSE must exhibit a global minimum for the fit to converge to a reasonable value for the SH losses, as is the case for the waveguide analysed in Figures \ref{fig:fitting_modes} and \ref{img:MSE}.%, e.g. in presence of noise or asymmetries in the phasematching spectrum that can arise from imperfect waveguides \cite{Santandrea19.NJP}, as these effects are not included in the model. 

However, a global minimum for the MSE was not always observed. This was seen when investigating a 10mm long waveguide from a second 7$\mu$m-wide titanium indiffused waveguide. The process under investigation was again a quasi-phase matched, type 0 second harmonic generation in the TM00 spatial mode with $\Lambda_0$=16.8 $\mu$m. 
This waveguide was found to have losses $\alpha_{\omega}$=0.12 $\pm$0.02dB/cm near to the phasematching wavelength of 1527nm. Using the method described in the previous section, the fit yielded vanishingly small (below 10$^{-4}$ dB/cm) second harmonic losses.%, since the MSE does not present a global minimum, as shown in figure \ref{mse}.

\begin{figure}[btp]
	\centering
	\includegraphics[width = 13cm]{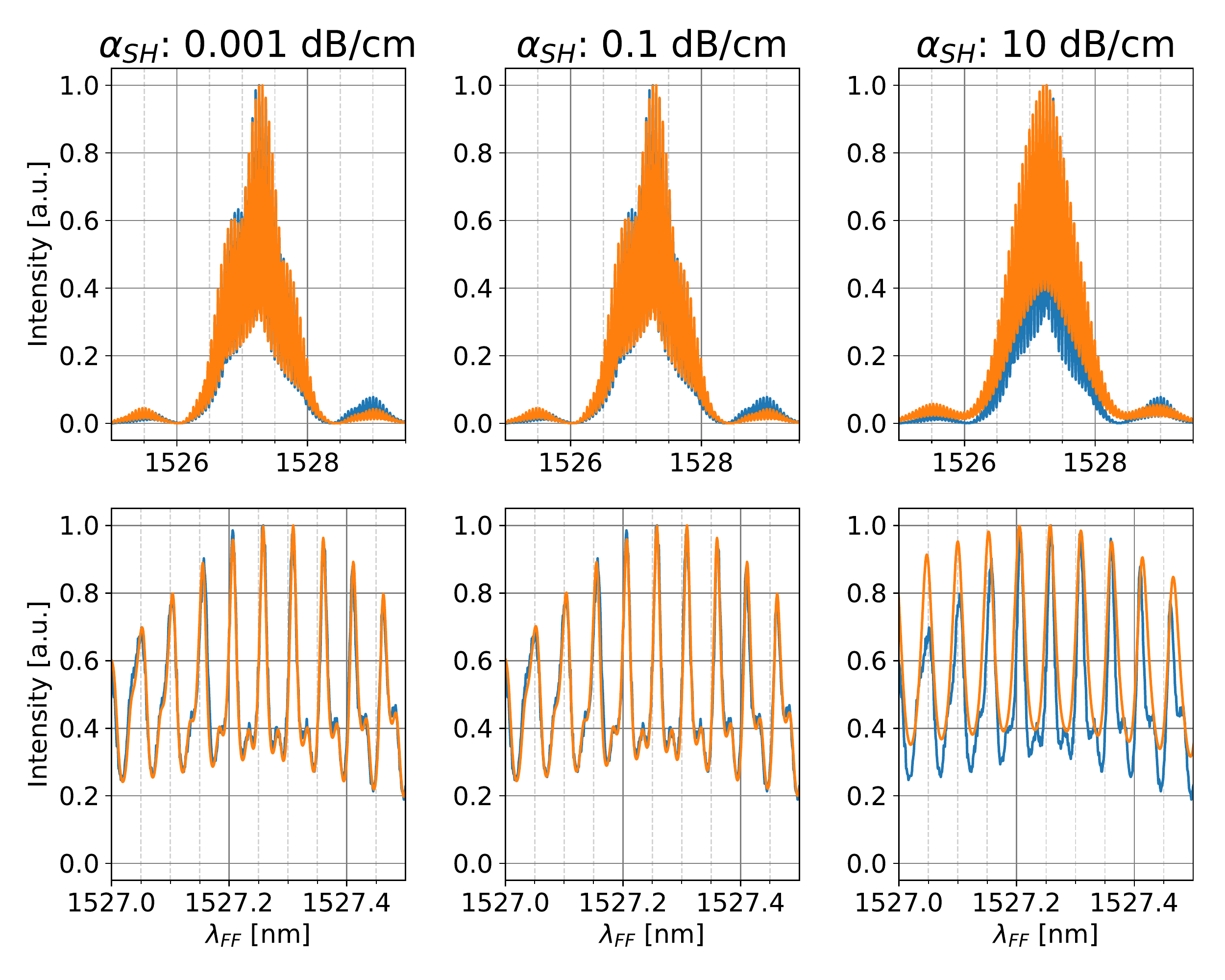}
	\caption{Central portion of the measured (blue) and fitted (orange) phasematching profiles. It can be qualitatively seen that the fit works very well for low losses but that both the structure and envelope of the fit for higher second harmonic losses is degraded. The bottom row shows a zoom-in on the region around 1527.2nm, highlighting the ability of the model to fit the fine structure of the measured spectrum.}
	\label{plots}
\end{figure}

\begin{figure}[btp]
	\centering
	\includegraphics[width = 10cm]{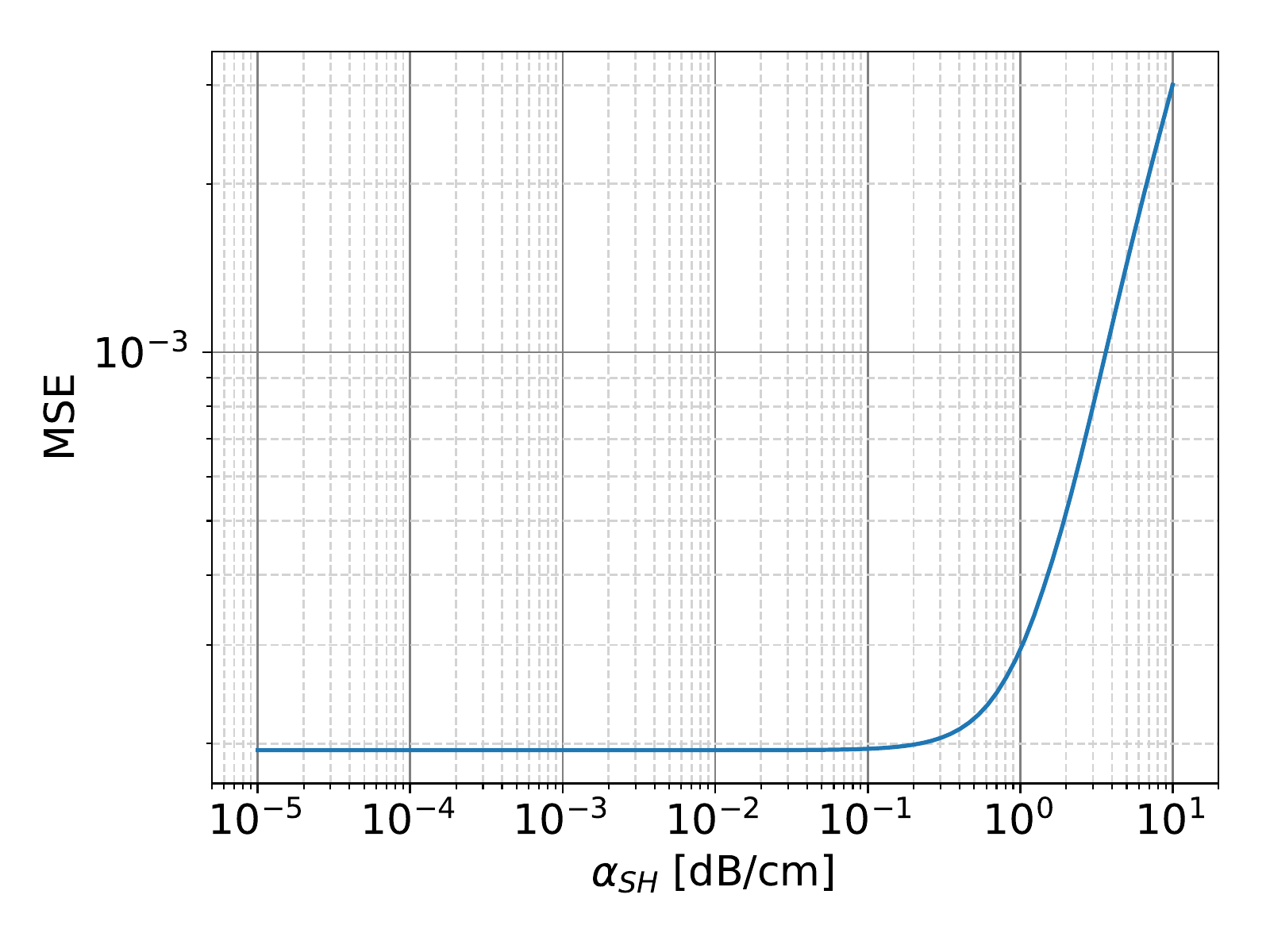}
	\caption{Mean squared error between the measured and fitted central portion of the phasematching profiles as the second harmonic losses are increased. It can be clearly seen that the mean squared error increases rapidly with losses greater than around 0.1 dB/cm. Of note is that the mean squared error does not increase with vanishingly small second harmonic losses.}
	\label{mse}
\end{figure}

However, as displayed in Figure \ref{plots}, a visual inspection of the fit for different values of $\alpha_{2\omega}$ reveals that, qualitatively, the measured spectrum can be reproduced very well with losses up to $\alpha_{2\omega}\lesssim$ 0.1 dB/cm. Furthermore, it can be seen that the fit noticeably degrades at the higher losses of 10dB/cm. 
The analysis of the MSE for varying second harmonic losses reveals the problem, as shown in Figure \ref{mse}. In contrast to the previously investigated waveguide, the MSE for this waveguide does not show a global minimum. In fact, the MSE asymptotes towards its lowest values at vanishingly small second harmonic losses. In this case, even though the minimisation routine fails to provide an estimate for the SH losses, the inspection of the MSE as a function of $\alpha_{2\omega}$ can still be used to determine an upper bound on the losses by choosing a threshold value in relation to the asymptote. For example, setting a 1\% threshold for the variation of the MSE with respect to its minimum value provides an upper bound of $\alpha_{2\omega}\leq$ 0.13 dB/cm for the second harmonic losses in this waveguide. This result and the previously determined losses for the first waveguide are summarised in Table \ref{tab: loss}.

\begin{table}[h!]
	\begin{center}
		\caption{Measured loss values of waveguide 1 (WG1) and waveguide 2 (WG2) for fundamental (TM00) and higher order (TM01) second harmonic spatial mode.}
		\label{tab:table1}
		\begin{tabular}{c|c|c|c} % <-- Alignments: 1st column left, 2nd middle and 3rd right, with vertical lines in between
			& \textbf{WG1 TM00} & \textbf{WG1 TM01} & \textbf{WG2 TM00}\\
			\hline
			Fund. Loss & 0.21$\pm$0.04dB/cm & 0.24$\pm$0.07dB/cm & 0.12$\pm$0.02dB/cm\\
			Harm. Loss & 1.2$\pm$0.2dB/cm & 1.3$\pm$0.1dB/cm & $\leq 0.13$dB/cm\\
		\end{tabular}
		\label{tab: loss}
	\end{center}
\end{table}

It is unclear why a global minimum for the MSE is not found in certain cases. It is likely that the chosen cost function is not sufficiently sensitive to small changes in the second harmonic losses, particularly in the presence of experimental imperfections. A more advanced fitting scheme may be able to predict the second harmonic losses with reduced uncertainty, but this is left as future work.

\section{Conclusion}

In this paper we have introduced a new method for characterizing the loss of spatially multi-mode waveguides. A model is introduced that describes the expected phasematching spectrum of the generated second harmonic power, including interferences due to the Fabry-Perot effect from the uncoated end facets. Experimental data is obtained by scanning the wavelength of the fundamental pump field over the phasematching spectrum corresponding to a chosen, single spatial-mode of the second harmonic field. In this way it is possible to determine the losses of a chosen spatial-mode of the second harmonic. The presented technique is then applied to two waveguides. In one case a reasonable estimate of the losses is found, and in the other an upper bound on the second harmonic losses is obtained. The presented approach is very general and can be extended to other nonlinear processes in virtually any high quality waveguide system.

\section*{Funding}
The work was supported by the European Union via the EU quantum flagship project UNIQORN (Grant No. 820474) and by the DFG (Deutsche Forschungsgemeinschaft).

%\bibliographystyle{unsrt}
%\bibliography{2019MarBib}

\begin{thebibliography}{10}

\bibitem{Stefszky17.PRA}
M.~Stefszky, R.~Ricken, C.~Eigner, V.~Quiring, H.~Herrmann, and C.~Silberhorn.
\newblock Waveguide cavity resonator as a source of optical squeezing.
\newblock {\em Phys. Rev. Applied}, 7:044026, 2017.

\bibitem{Stefszky17Comb.A}
M.~{Stefszky}, V.~{Ulvila}, C.~{Silberhorn}, and M.~{Vainio}.
\newblock {Towards optical frequency comb generation in continuous-wave pumped
  titanium indiffused lithium niobate waveguide resonators}.
\newblock {\em ArXiv e-prints}, December 2017.

\bibitem{Kaiser16.O}
F.~Kaiser, B.~Fedrici, A.~Zavatta, V.~D'Auria, and S.~Tanzilli.
\newblock A fully guided-wave squeezing experiment for fiber quantum networks.
\newblock {\em Optica}, 3(4):362--365, Apr 2016.

\bibitem{Barral17.PRA}
David Barral, Nadia Belabas, Lorenzo~M. Procopio, Virginia D'Auria, S\'ebastien
  Tanzilli, Kamel Bencheikh, and Juan~Ariel Levenson.
\newblock Continuous-variable entanglement of two bright coherent states that
  never interacted.
\newblock {\em Phys. Rev. A}, 96:053822, Nov 2017.

\bibitem{Regener85.APB}
R.~Regener and W.~Sohler.
\newblock Loss in low-finesse {Ti:LiNbO$_{3}$} optical waveguide resonators.
\newblock {\em Appl. Phys. B}, 36:143, 1985.

\bibitem{Park95.JAP}
K.~H. Park, M.~W. Kim, Y.~T. Byun, D.~Woo, S.~H. Kim, S.~S. Choi, Y.~Chung,
  W.~R. Cho, S.~H. Park, and U.~Kim.
\newblock Nondestructive propagation loss and facet reflectance measurments of
  {GaAs/AlGaAs} strip-loaded waveguides.
\newblock {\em J. Appl. Phys.}, 78:6318, 1995.

\bibitem{Clark90.OL}
D.F. Clark and M.~S. Iqbal.
\newblock Simple extension to the fabry-perot technique for accurate
  measurement of losses in semiconductor waveguides.
\newblock {\em Opt. Lett.}, 15:1291, 1990.

\bibitem{DeRossi05.JAP}
Alfredo De~Rossi, Valentin Ortiz, Michel Calligaro, Loïc Lanco, Sara Ducci,
  Vincent Berger, and Isabelle Sagnes.
\newblock Measuring propagation loss in a multimode semiconductor waveguide.
\newblock {\em Journal of Applied Physics}, 97(7):073105, 2005.

\bibitem{Ruetz16.APB}
H.~R\"{u}tz, K-H. Luo, H.~Suche, and C.~Silberhorn.
\newblock Towards a quantum interface between telecommunication and {UV}
  wavelengths: {D}esign and classical performance.
\newblock {\em Appl. Phys. B}, 122:13, 2016.

\bibitem{Berger97.JOSAB}
V.~Berger.
\newblock Second-harmonic generation in monolithic cavities.
\newblock {\em J. Opt. Soc. Am. B}, 14(6):1351--1360, Jun 1997.

\bibitem{Fujimura96.JLT}
M.~{Fujimura}, T.~{Suhara}, and H.~{Nishihara}.
\newblock Theoretical analysis of resonant waveguide optical second harmonic
  generation devices.
\newblock {\em Journal of Lightwave Technology}, 14(8):1899--1906, Aug 1996.

\bibitem{Santandrea19.NJP}
M~Santandrea, M~Stefszky, V~Ansari, and C~Silberhorn.
\newblock Fabrication limits of waveguides in nonlinear crystals and their
  impact on quantum optics applications.
\newblock {\em New Journal of Physics}, 2019.

\bibitem{Strake1988}
E.~Strake, G.~P. Bava, and I.~Montrosset.
\newblock {Guided Modes of Ti:LiNbO3 Channel Waveguides: A Novel
  Quasi-Analytical Technique in Comparison with the Scalar Finite-Element
  Method}.
\newblock {\em Journal of Lightwave Technology}, 6(6):1126--1135, 1988.

\bibitem{Jundt1997}
D.H. Jundt.
\newblock {Temperature-dependent Sellmeier equation for the index of
  refraction, n(e), in congruent lithium niobate.}
\newblock {\em Optics Letters}, 22(20):1553--1555, 1997.

\end{thebibliography}

\end{document}